\documentclass[preprint, amsmath,amssymb]{revtex4}
\usepackage{graphicx}
\begin{document}
%\section{Comment on ``Parametric amplification in Josephson junction embedded transmission lines''}
\title{Comment on ``Parametric amplification in Josephson junction embedded transmission lines''}
\begin{abstract}

%\subsection{Abstract}

Recently Yaakobi and co-workers [Phys. Rev. B \textbf{87}, 144301 (2013)] theoretically studied four-wave mixing and parametric amplification in a nonlinear transmission line consisting of capacitively shunted Josephson junctions. By deriving and solving the coupled-mode equations, they have arrived at the conclusion that in a wide frequency range around the pump frequency exponential parametric gain (in which the signal grows exponentially with distance) can be achieved. However, we have found a mathematical error in their derivation of the coupled-mode equations (Equation (A13)), which leads to the wrong expression of the gain factor and invalidates their conclusions on the gain and bandwidth. In this comment, we present the correct expression for the parametric gain. We show that for a transmission line with weak dispersion or positive dispersion ($\Delta k>0 $), as is the case discussed by Yaakobi et al, while quadratic (power) gain can occur around the pump frequency, exponential gain is impossible. Furthermore, for a transmission line with proper intrinsic or engineered dispersion, exponential gain occurs at frequencies where the phase matching condition is met, while around the pump frequency the gain is still quadratic.

\end{abstract}

\author{S. Chaudhuri}
\affiliation{Stanford University, Department of Physics, Stanford, CA 94305}

\author{J. Gao}
\affiliation{National Institute of Standards and Technology, Boulder, CO 80305}

\date{\today}

\maketitle

%\subsection{Text}

Superconducting parametric amplifiers are currently of great interest because of their promise in achieving quantum-limited noise, which has important applications in the readout of superconducting quantum bits \cite{siddiqi} and astronomical low-temperature detectors. \cite{day} In a recent paper \cite{yaakobi} (referred to as ``the Paper'' hereafter), Yaakobi and co-workers theoretically studied four-wave mixing and parametric amplification in a nonlinear transmission line comprised of capacitively shunted Josephson junctions. By deriving and solving the coupled mode equations, they have derived an exponential gain factor g (signal grows exponentially with the distance, $y_s(x) \sim e^{gx}$). 
%Under the long wavelength approximation (frequency far below plasma frequency of the junctions), 
The final expression for g, implied by the Paper, is 

\renewcommand \theequation{\Roman{equation}}
\begin{equation}
g = \sqrt{-\Delta_p}=\sqrt{\left(\frac{3\gamma \tilde{k}_{p} ( \tilde{k}_{s} \tilde{k}_{i})^{1/2}}{4} B_{p,0}^{2} \right)^{2} - \left(\frac{3\gamma \tilde{k}_{p}^{2}}{8} B_{p,0}^{2} \left( 1+ \frac{\Delta \tilde{k}}{\tilde{k}_{p}} \right) + \frac{\Delta k}{2} \right)^{2} }
\label{eqn:g_wrong}
\end{equation}
which reaches a maximum positive value of $g_\mathrm{max}$ at the pump frequency (Eqn. (57)) \footnote{Throughout this comment, equations labeled with arabic numbers and roman numerals refer to equations defined in the original paper and this comment, respectively. } and remains positive in a bandwidth $\Delta \omega_B$ (Eqn.~(56)). Yaakobi et al have drawn the conclusion that this traveling wave parametric amplifier architecture can generate exponential gain in a frequency range around the pump frequency. In this comment, we show that this conclusion is invalid, due to an error in their derivation of the gain factor.

In careful examination of their appendix, we have found a mathematical error in the derivation of the coupled-mode equations. In the cubic expansion of the mixing products, the coefficients in front of the terms $k_p^2k_iA_p^2A_i^*e^{-i\Psi}$ and $k_p^2k_sA_p^2A_s^*e^{-i\Psi}$ in Eqn. (A13) should be 1 instead of 2. This leads to a factor of 2 reduction in $\mu$ in Eqn.~(31) and a different value of $g$. The discussions on the gain and bandwidth, based on the wrong expression of $g$, Eqn.~\ref{eqn:g_wrong}, are therefore invalid. Here we present the correct expression for $g$,

\begin{equation}
g = \sqrt{-\Delta_p}=\sqrt{\left(\frac{3\gamma \tilde{k}_{p} ( \tilde{k}_{s} \tilde{k}_{i})^{1/2}}{8} B_{p,0}^{2} \right)^{2} - \left(\frac{3\gamma \tilde{k}_{p}^{2}}{8} B_{p,0}^{2} \left( 1+ \frac{\Delta \tilde{k}}{\tilde{k}_{p}} \right) + \frac{\Delta k}{2} \right)^{2} }
\label{eqn:g_correct}
\end{equation}

%In the long-wavelength limit, which is the case discussed in Yaakobi et al, $\tilde{k}_{m} \approx k_{m} %\approx \omega_{m}$ and $\Delta \tilde{k} \approx 9\tilde{k}_{p}(\Delta\omega)^{2} \ll \tilde{k}_{p}$ (from %Eqn. (A32)) ***(confirm that the second approximation is okay to use)*** and the above result becomes
%\begin{equation}
%g = \sqrt{\frac{\omega_{s}\omega_{i}}{\omega_{p}^{2}} \left(\frac{3\gamma \tilde{k}_{p}^{2}}{8} B_{p,0}^{2} %\right)^{2} - \left(\frac{3\gamma \tilde{k}_{p}^{2}}{8} B_{p,0}^{2}  + \frac{\Delta k}{2} \right)^{2} }
%
%\end{equation}

To make a connection with established fiber parametric amplifier theory \cite{agrawal}, we write the exponential gain factor $g$ and the signal power gain $G_{s}$ in a more general form,
\begin{equation}
g = \sqrt{-\Delta_{p}}=\sqrt{\frac{\omega_{s}\omega_{i}}{\omega_{p}^{2}} \left( \Delta \Phi_4 \right)^{2}- \left( \frac{\kappa_{4}}{2} \right)^{2}}
\label{eqn:g_sc}
\end{equation}
\begin{equation}
G_s = \left| \mathrm{cosh}(gL) + \frac{i\kappa_{4}}{2g} \mathrm{sinh}(gL) \right|^{2},\\
\label{eqn:Gs_sc}
\end{equation}
where $\Delta \Phi_4$ is the pump self-phase modulation per unit length, $\kappa_{4}=\Delta k + 2\Delta\Phi_4$ is a measure of the phase mismatch, and $L$ is the length of the traveling-wave structure. For the Josephson junction transmission line discussed in the Paper, $\Delta \Phi_4 = \frac{3\gamma \tilde{k}_{p}^{2}}{8} B_{p,0}^{2}$. The expression Eqn.~\ref{eqn:g_sc} matches the corrected $g$ value in Eqn.~\ref{eqn:g_correct} in the long-wavelength limit ($\omega_{m} \propto \tilde{k}_{m}$ and $\Delta \tilde{k}/\tilde{k}_p \ll 1$), and Eqn.~\ref{eqn:Gs_sc} matches Eqn. (51) in the Paper. The expressions for $g$ and $G_s$ also apply to other four-wave mixing traveling wave parametric amplifiers whose phase velocity exhibits quadratic nonlinearity with respect to the wave amplitude, such as the dispersion-engineered traveling-wave kinetic inductance (DTWKI) amplifier \cite{eom}.

\begin{figure}[ht]
  \centering
  \includegraphics[width=\textwidth]{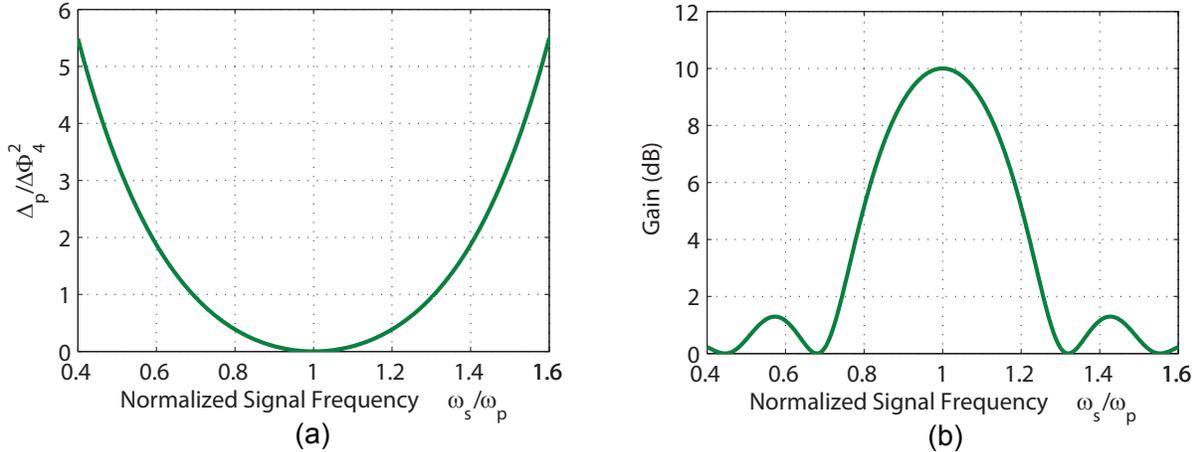}
  \caption{$\Delta_p$ and $G_s$ as a function of signal frequency from Eqn.~\ref{eqn:g_sc} and \ref{eqn:Gs_sc}. The parameters used to generate these curves are from page 5 of the Paper. The total self-phase shift is set to $L\Delta\Phi_{4}=3$ radians, resulting in a maximum gain of $G_{s,max}=1+(L\Delta\Phi_{4})^{2}=10$ dB (from Eqn.~\ref{eqn:Gs_linear}).}
  \label{fig1}
\end{figure}

The correct gain expressions show significantly different features from those discussed in the Paper. A direct implication from Eqn.~\ref{eqn:g_sc} is that $\Delta_p$ has to be negative in order for exponential gain to occur, which in turn requires $\Delta k < 0$. From Eqn.~(A32) we see that $\Delta k \geq 0$ for the transmission line discussed in the Paper. This can also be seen from a non-negative $\Delta_p$ in Fig.~\ref{fig1}(a), where $\Delta_p$ is plotted for the set parameters given on page 5 of the Paper. Therefore, exponential gain is impossible at any frequency for this transmission line. Also, the gain coefficient g is always imaginary. 

Furthermore, the transmission line discussed in the Paper has dispersion coming from the ladder network and the Josephson capcitance ($C_J$). The amount of dispersion is small near the pump frequency where $\Delta k \ll 2\Delta \Phi_4$ holds.
%in the assumed long wavelength limit, $\omega_m \ll \omega_p \ll \omega_c$, where $\omega_p$ and $\omega_c$ are the %plasma frequency of the junction and the cutoff frequency of the laddered line, respectively. For a realistic %device illustrated by Yaakobi and et al
For a transmission line with no dispersion ($\Delta k =0$) or with weak dispersion ($|\Delta k| \ll 2\Delta \Phi_4$), Eqn.~\ref{eqn:g_sc} and Eqn.~\ref{eqn:Gs_sc} further reduce to
\begin{equation}
\Delta_p = \left( 1-\frac{\omega_s}{\omega_p} \right)^2\Delta \Phi_4^2,~~ g =\sqrt{-\Delta_p}= i \left|1-\frac{\omega_{s}}{\omega_{p}} \right| \Delta\Phi_{4}
\label{eqn:g_linear}
\end{equation}
\begin{equation}
G_s = 1+ \left( \frac{1}{\left| 1- \frac{\omega_{s}}{\omega_{p}} \right|^{2}}-1 \right) \mathrm{sin}^{2} \left( \left| 1-\frac{\omega_{s}}{\omega_{p}} \right| L \Delta\Phi_{4} \right)
\label{eqn:Gs_linear}
\end{equation}
 %First, $\Delta_p$ is non-negative at all signal frequencies (see Fig. 1a, the value $\Delta_{p}/\Delta\Phi_{4}^{2}$ vs signal frequency), and consequently, the gain coefficient g is imaginary. 
As $\omega_{s} \rightarrow \omega_{p}$, $g\rightarrow 0$ and the signal power gain $G_s$ approaches a maximum of $G_{s,max}=1+(L\Delta\Phi_{4})^{2}$. The power gain around the pump frequency becomes quadratic (signal amplitude grows linearly with distance). Therefore we conclude that, \textbf{for the case considered by Yaakobi et al, while quadratic gain can occur around the pump frequency, exponential gain is impossible at all frequencies.} 

For a transmission line with proper dispersion, $|\Delta k| \sim 2\Delta\Phi_{4}$ and $\Delta k < 0$ at some frequency, exponential gain can occur. When the phase-matching condition $\kappa_4 \simeq 0$ is met or nearly met, $g$ can become real and positive. However, since $\Delta k \simeq 0$ for signal frequencies near the pump, as long as $k$ is a continous function of $\omega$, the gain around the pump will still be quadratic. These features have been verified in the measured gain profiles from both the optical fiber parametric amplifier (with intrinsic dispersion) and the DTWKI amplifier (with engineered dispersion), where one can identify exponential gain regions in two broader frequency bands detuned from the pump frequency and a quadratic gain region in a narrower band around the pump \cite{agrawal,eom}. 

We want to point out that the Josephson junction transmission line considered by Yaakobi et al can still operate as a parametric amplifer. However, the gain and bandwidth are much reduced from those discussed in the Paper. Plotted in Fig.~\ref{fig1}(b) is the gain profile for a total self-phase shift of $L\Delta\Phi_{4}=3$ radians for the parameters given on pg. 5 of the Paper. Now the 3-dB bandwidth reduces to $\sim 30\%$ and the maximum gain is 10~dB. Because the maximum gain is limited to be quadratic, in order to achieve a certain gain, $L\Delta\Phi_{4}$ has to be larger than the exponential case discussed in the Paper. This means a longer line (more junctions) or driving the junctions closer to their critical current $I_c$. The amplifier presented in Fig.~\ref{fig1}(b) could potentially give a maximum exponential gain of $G_s = 20$~dB, according to Eqn.~\ref{eqn:Gs_sc}, and a broader bandwidth, if proper dispersion is introduced and the phase-matching condition is met. Therefore, it is very interesting to consider applying the dispersion engineering concept used in the DTWKI amplifer to the Josephson junction traveling wave parametric amplifier.


\begin{thebibliography}{9}
\bibitem{siddiqi}
R. Vijay, C. Macklin, D.H. Slichter, S.J. Weber, K.M. Murch, R. Naik, A. Korotkov, I. Siddiqi, Nature, \textbf{490}, 77 (2012).
\bibitem{day}
P. K. Day, H.G. Leduc, B.A. Mazin, A. Vayonakis, and J. Zmuidzinas, Nature, \textbf{425}, 817 (2003).
\bibitem{yaakobi}
O. Yaakobi, L. Friedland, C. Macklin, and I. Siddiqi, Phys. Rev. B, \textbf{87}, 144301 (2013).
%\bibitem{Chaudhuri}
%S. Chaudhuri, et al. In preparation.
\bibitem{agrawal}
G.P. Agrawal, \emph{Nonlinear Fiber Optics}, 3rd ed. (Academic Press, San Diego, 2001).
\bibitem{eom}
B. Ho Eom, P.K. Day, H.G. Leduc, and J. Zmuidzinas, Nat. Phys. \textbf{8}, 623 (2012).
\end{thebibliography}
\end{document}